# A practical efficient and effective method for the Hamiltonian cycle problem that runs on a standard computer

Eric Lewin Altschuler*, Timothy J. Williams**

*Department of Physical Medicine and Rehabilitation Temple University School of Medicine, 3401 N. Broad St., Philadelphia, PA 19140, USA

**Leadership Computing Facility, Argonne National Laboratory 9700 S. Cass Ave, Bldg. 240, Argonne, IL 60439, USA

Abstract

Given $N$ cities and $R < N^2-N$ directed (unidirectional/one way) roads does there exist a tour of all $N$ cities stopping at each city exactly once using the given roads (a Hamiltonian cycle)? This Hamiltonian cycle problem (HCP) is an NP-complete problem, for which there is no known polynomial time solution algorithm. The HCP has important practical applications, for example, to logistical problems. It was claimed that an adiabatic quantum computer could solve an NP-complete problem faster than classical algorithms, but claim appears to have been debunked. Here we demonstrate an algorithm which runs on a standard computer that efficiently and effectively solves the HCP for at least up to 500 cities: We first optimized a simulated annealing based algorithm used for smaller sized HCP problems. Then we found that when a tour was deliberately inserted in a list of otherwise randomly chosen roads, crucially, if "extra" random roads are added to bring the total number of roads up to $0.58*N*\log_e N$ or more there is a 100% chance our algorithm will find a HC, but conversely when a list of roads does not include a pre-inserted tour random roads have to be added until there are $0.9*N*\log_e N$ roads to have a chance of finding a HC. We found similarly for a set of roads non-randomly chosen. Thus, the presence of a HC in a set of roads induces "connectivity" throughout the roads and a HC can be found with an insertion of a modest number of extra roads. Our algorithm also shows that only weakly non-local information is needed to find an HCP that is a global state.

Given $N$ cities and $R < N^2-N$ directed (unidirectional/one way) roads does there exist a tour of all $N$ cities stopping at each city exactly once using the given roads (a Hamiltonian cycle)? This Hamiltonian cycle problem (HCP) is an NP-complete problem, for which there is no known polynomial time solution algorithm.[1] The HCP has important practical applications, for example, to logistical problems. While there was a claim[2] that an adiabatic quantum computer could solve an NP-complete problem faster than classical algorithms, this claim appears to have been debunked[3]. Here we demonstrate an algorithm which runs on a standard computer that efficiently and effectively solves the HCP for at least up to 500 cities: We first optimized a simulated annealing[4] based algorithm[5] used for smaller sized HCP problems. Then we found that when a tour was deliberately inserted in a list of otherwise randomly chosen roads, crucially, if "extra" random roads are added to bring the total number of roads up to $0.58*N*\log_e N$ or more there is a 100% chance our algorithm will find a HC, but conversely when a list of roads does not include a pre-inserted tour random roads have to be added until there are $0.9*N*\log_e N$ roads to have a chance of finding a HC. We found similarly for a set of roads non-randomly chosen. Thus, the presence of a HC in a set of roads induces "connectivity" throughout the roads and a HC can be found with an insertion of a modest number of extra roads. Our algorithm also shows that only weakly non-local information is needed to find a HC that is a global state.

Simulated annealing[4] is an approach inspired by statistical mechanics that by analogy views the values of a multivariable numeric problem as physical states of particles. Simulated annealing is very useful practically for problems such as the traveling salesperson problem and other so called NP-complete problems that have no known polynomial time numeric solutions.

To employ simulated annealing one needs first to define a "cost" or energy function and also "moves" to perturb the state of the system. We take a tour as a permutation of the numbers 1,2,3…$N$. The tour length is the cost function we use. We assign the distance between two cities 0 if there exists a road between the cities, and 1 if not. A tour of length 0 thus corresponds to a Hamiltonian cycle. When perturbing the energy function in using simulated annealing ideally one wants to keep much of a configuration the same and only change a small portion of the configuration. The idea is to retain the "good" parts of the configuration and find and modify the high-energy parts. If a tour is not a Hamiltonian cycle we make a new tour randomly using one of two "moves" to modify the tour: The first move we use is the "transport" move of Lin and Kernighan[6] (Figure 1a) which transports a subsegment of the tour to a new location in the tour. We do not use the "reversal" move of Lin and Kernighan[6], because, as shown in Figure 1b, this would change the direction of travel between potentially many cities, but since the roads are one way a road between the same cities but in the opposite direction might not be in the given set of roads so cannot be used to form a Hamiltonian cycle. Instead of the reversal move we use a "swap" move[5] (Figure 1c) that interchanges the positions of two cities in the tour. At most this affects only four directional roads and thus keeps more of the original tour intact than the reversal move. We now compute the length of the new tour. If the length of the new tour is shorter than the prior tour, it is accepted as the new tour. Even if the new tour is longer, we occasionally accept the longer tour using the Metropolis algorithm.[7]

(a)

|     |    |    |    |    |    |    |    |    |    |     |
|-----|----|----|----|----|----|----|----|----|----|-----|
| ... | 37 | 8  | 42 | 90 | 27 | 23 | 65 | 13 | 2  | 55 ... |
| ... | 37 | 8  | *42* | *90* | *27* | 23 | 65 | 13 | 2  | 55 ... |
| ... | 37 | 8  | 42 | 90 | 27 | 23 | 65 | *13* | 2  | 55 ... |
| ... | 37 | 8  | 23 | 65 | *13* | *42* | *90* | *27* | 2  | 55 ... |

(b)

|     |    |    |    |    |    |    |    |    |    |     |
|-----|----|----|----|----|----|----|----|----|----|-----|
| ... | 37 | 8  | 42 | 90 | 27 | 23 | 65 | 13 | 2  | 55 ... |
| ... | 37 | 8  | *42* | *90* | *27* | *23* | 65 | 13 | 2  | 55 ... |
| ... | 37 | 8  | *23* | *27* | *90* | *42* | 65 | 13 | 2  | 55 ... |

(c)

|     |    |    |    |    |    |    |    |    |    |     |
|-----|----|----|----|----|----|----|----|----|----|-----|
| ... | 37 | 8  | 42 | 90 | 27 | 23 | 65 | 13 | 2  | 55 ... |
| ... | 37 | 8  | *42* | 90 | 27 | *23* | 65 | 13 | 2  | 55 ... |
| ... | 37 | 8  | *23* | 90 | 27 | *42* | 65 | 13 | 2  | 55 ... |

Figure 1: System perturbing moves for simulated annealing algorithm for the HCP. **a,** Lin-Kernighan "transport" move.[6] **b,** Lin-Kernighan "reversal" move.[6] **c,** "Swap" move[5].

At each annealing step, try some number of configurations—rearrangements of the order of cities in the tour, using the moves detailed above to generate the configurations. The chance of accepting a configuration resulting in a longer tour is reduced as time goes on; by analogy, the problem cools according to an annealing schedule. From an initial "temperature" value $T_0$, given a cooling factor $F_c$, we iterate through $N_a$ annealing steps, each time reducing the temperature by the cooling factor. That is, moving from step $i$ to step $i+1$,

$$T_{\{i+1\}} = T_i * F_c$$

The Metropolis condition is

$$\exp(-|\delta L|/kT)$$

where $\delta L$ is the increase/decrease in tour length after a given move, and $k$ is a Boltzmann-like constant. The number of configurations tried in each annealing step is equal to $MN^2$, where $M$ is a multiplicative constant—a tunable parameter. The roads connecting pairs of cities are initialized randomly, and there are $mN \log_e N$, where $m$ is a tunable multiplicative constant.

Certainly, if a set of roads does not contain a Hamiltonian cycle our algorithm will correctly report that there is no HC. But since it cannot be known a priori (for any sizeable set of roads, e.g., 40 cities) that the roads do not contain a HC to study an algorithm one needs to use a set of roads into which a HC has been placed and see if the algorithm will report that there is a HC, additionally one can see if the algorithm has found the "planted" HC or another one.

We first optimized an algorithm[5] for the HCP that was effective in finding HCs for $N$ up to 50 cities but then began to become significantly less effective. We took the case of $N = 150$ and with $0.5*150*\log_e 150$ roads as this number of roads for a given $N$ seems to be theoretically the most difficult case.[8] This extreme difficulty of the HCP around $0.5N \log_e N$ roads seems related though not exactly the same as phase transition like behavior Kirkpatrick and Selman found[9] in studying the k-satisfiability (k-SAT) problem: given a randomly generated Boolean expression of N (Boolean) variables built of clauses having k variables per clause, can the expression be satisfied (made true) with a single combination of true/false values for all the variables. For k=2, k-SAT has a polynomial time solution but for k >2 the problem is known to be NP-complete. It was found that for random $(2 + \varepsilon)$-SAT problems the ability to find solutions numerically the probability of finding a solution goes through a phase transition.

The Boltzmann constant $k$ changes the threshold for accepting "bad" moves, which increase the length of the trip rather than decrease it (that is, they increase the number of nodes in the current tour with no graph edge connecting them). Smaller values of $k$ decrease the threshold, so that "bad" moves are less frequent. Figure 2 shows the effect of $k$, and the "sweet spot" at $k=0.4$ maximizing the number of Hamiltonian cycles found out of 128 trials.

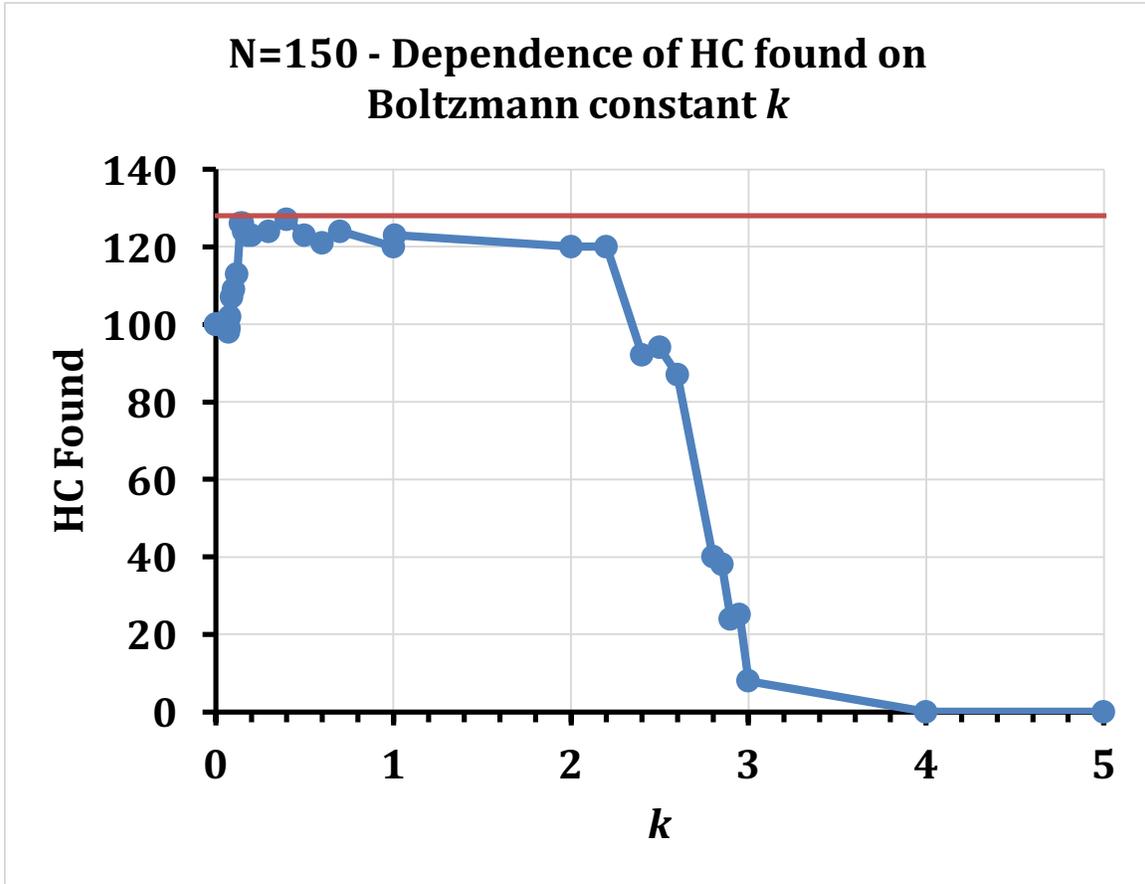

Figure 2 Dependence of the number of HCs found on *k*. Horizontal line is 128, the maximum possible.

Next using this optimized value for *k* we evaluated the algorithm starting with varying numbers of random roads. The number of roads is $m*N*\log_e N$, where *m* is a multiplicative constant. Results are shown in Figure 3. We see that a HC is found in all cases for $0.58*N*\log_e N$ or more roads. We also see that for *m*=0.5 or greater the inserted tour is never found and that for *m*=0.54 or greater all 128 HCs found are unique.

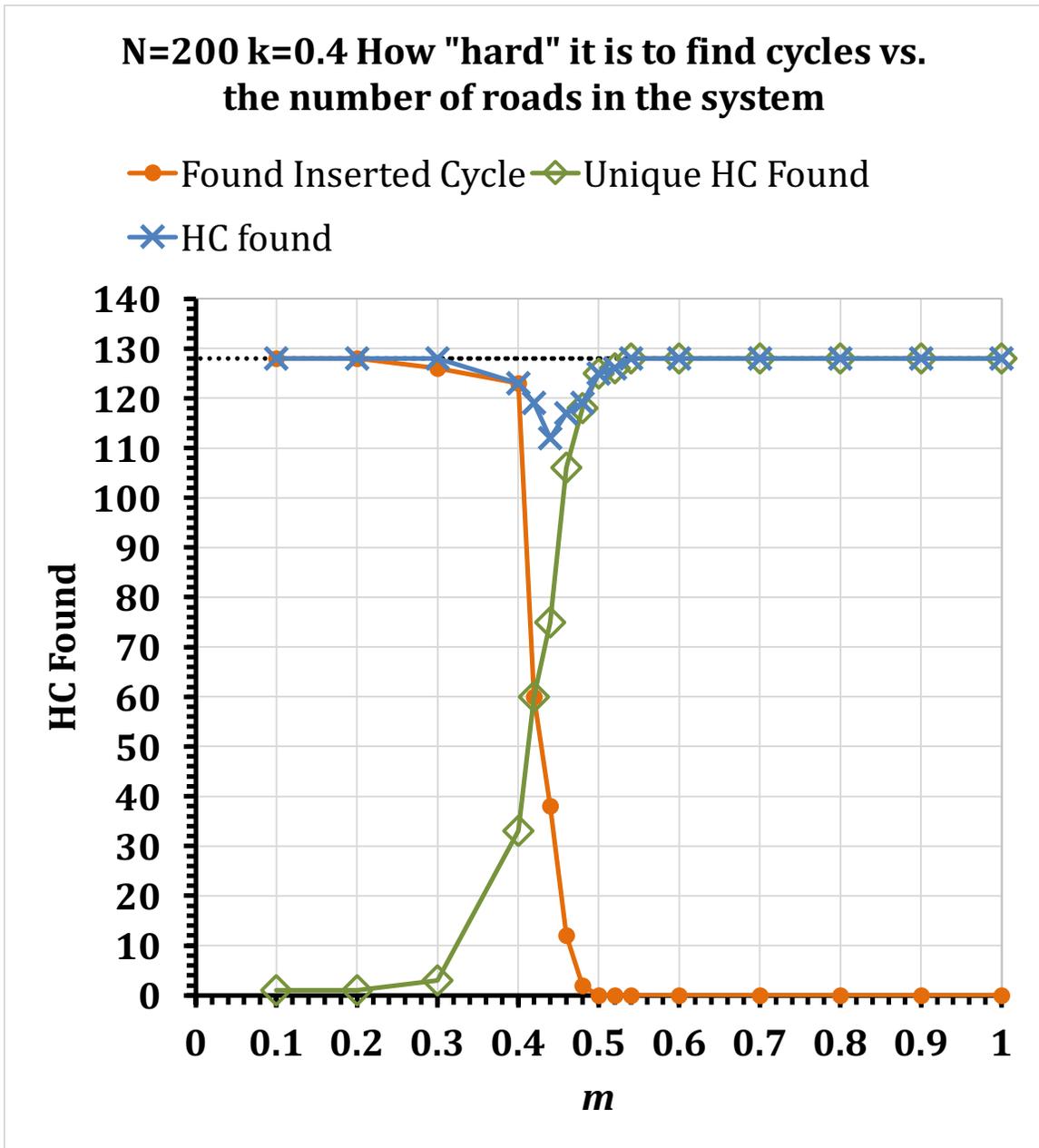

Figure 3 Probability of finding a HC as a function of number of roads. The number of roads is $m*N \log_e N$. The dotted horizontal line is 128, the maximum possible value.

We next asked what would happen if we did not insert a tour. As shown in Figure 4 for $N$=200, no HC was found until there were $0.9*N*\log_e N$ total roads. Thus the presence of even one HC (e.g., an inserted tour) seems to induce sufficient "connectivity" in the system so as to generate more HCs.

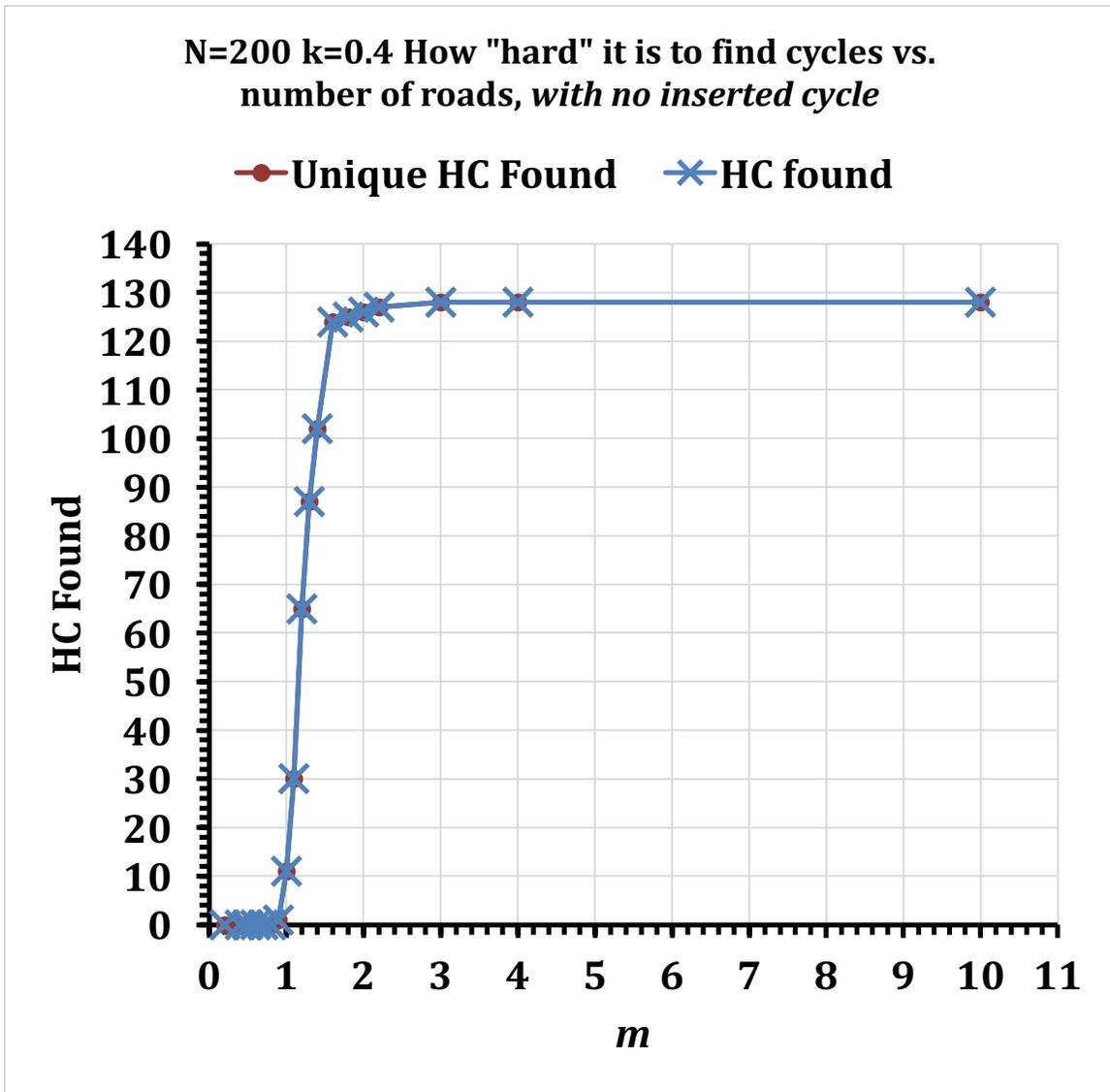

Figure 4 Probability of finding a HC as a function of number of roads when there is no pre-inserted HC. The number of roads is $m*N \log_e N$. At $m=4$ and higher, 128 HCs are found, the maximum possible number.

What about systems in which the edges are nonrandom? We ran some experiments with $MN^2$ deterministic edges in the graph, guaranteeing no Hamiltonian cycle. We generate the edges starting with stride 1, connecting every other pair of nodes; the sequence is {0→1, 2→3, 3→4, 5→6,….} until we reach node $N$ (or $N$-1, depending on whether $N$ is even or odd), or until we have the desired number of edges. If stride-1 reaches $N$ before we have the specified number of edges, we move on to stride-2: {0→2, 3→5, 6→8, 9→11,….}. If we have not hit the quota of edges yet within this sequence, we move on to stride-3: {0→3, 4→7, 8→11, 12→15,….}. We continue on up through larger strides as far as needed to generate the specified number of edges. To this we add, as before, one random-order Hamiltonian cycle. Results are shown in Figure 5. As for the case of roads added randomly to a HC, when there are $0.58*N*\log_e N$ total roads a HC is always found.

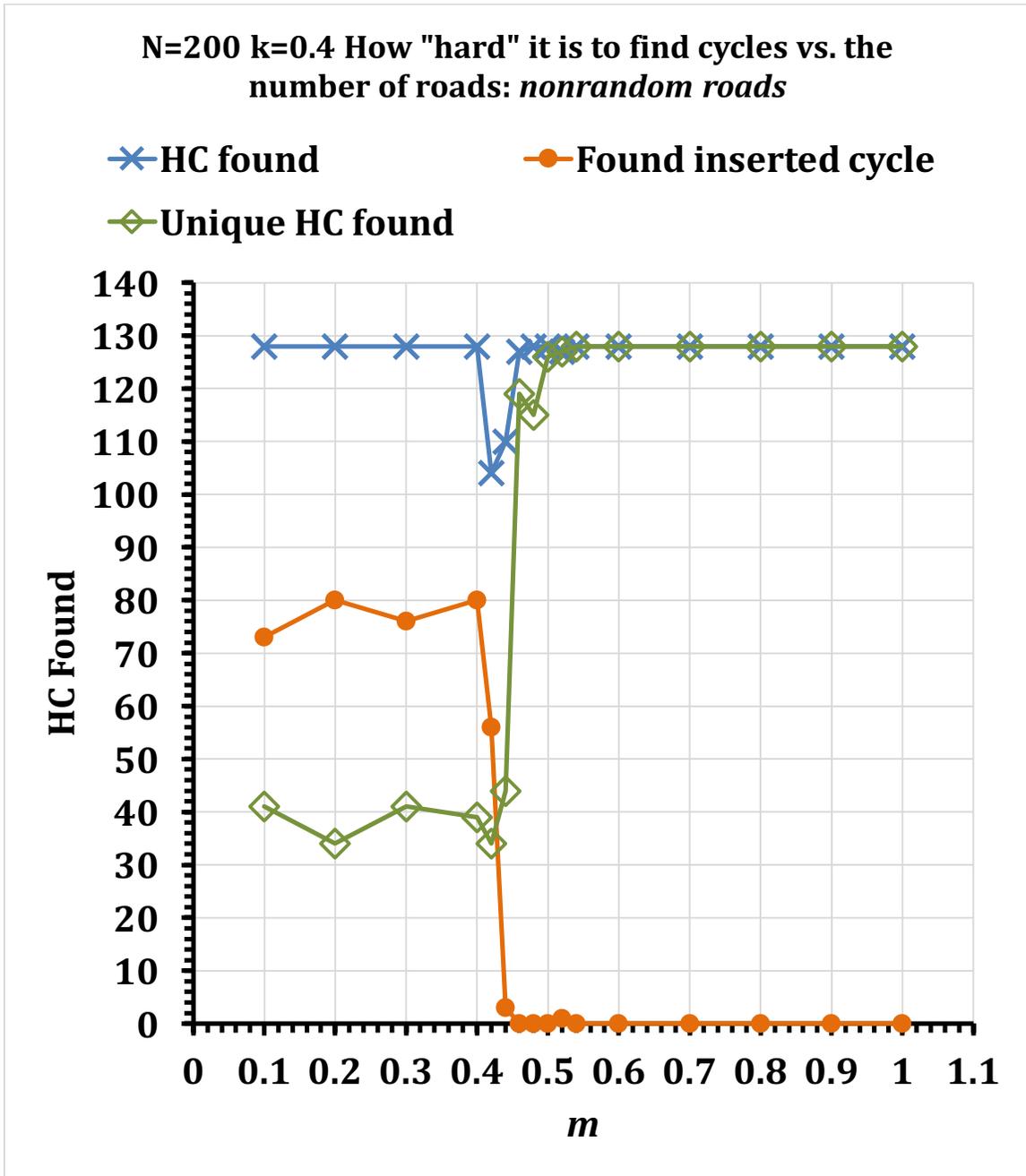

Figure 5 Probability of finding a HC as a function of number of roads when there is an inserted HC and other roads are added in a non-random fashion. The number of roads is $m*N \log_e N$. See text for method of adding roads non-randomly.

Again, as in the case when a collection of roads is generated randomly without an inserted HC, for a non-randomly generated set of even $0.9*N*\log_e N$ roads (in the same fashion as above) no HC is found (Table 1).

| m | HC found |
|---:|---:|
| 0.1 | 0 |
| 0.2 | 0 |
| 0.3 | 0 |
| 0.4 | 0 |
| 0.42 | 0 |
| 0.44 | 0 |
| 0.46 | 0 |
| 0.48 | 0 |
| 0.5 | 0 |
| 0.52 | 0 |
| 0.54 | 0 |
| 0.6 | 0 |
| 0.7 | 0 |
| 0.8 | 0 |
| 0.9 | 0 |
| 1 | 0 |

Table 1 **Probability of finding an HC as a function of number of roads for a set of non-randomly generated roads, with no cycle inserted. The number of roads is $m*N \log_e N$. Because of the way the non-random roads are generated, they are guaranteed not to contain an HC.**

We thus suggest the following practical algorithm for the HCP: To the given set of roads add randomly generated roads until there are at least $0.58*N*\log_e N$ total roads and run our annealing based algorithm to look for a HC. Given the results in Figure 3 and Figure 4, if a HC is found then there is an HC in the original set of roads. If a HC is not found, the original set of roads does not contain an HC. To test this we tried the cases of $N = 300, 400$ and $500$. Results are shown in Table 2.

| N | M | Cycles Found |
|---|---|---|
| 200 | 0.5 | 125 |
| 200 | 0.7 | 128 |
| 300 | 0.5 | 125 |
| 300 | 0.7 | 128 |
| 400 | 0.5 | 124 |
| 400 | 0.7 | 128 |
| 500 | 0.5 | 126 |
| 500 | 0.7 | 128 |

Table 2 **Probability of finding and HC for large *N* problems. In all cases, all the HC found are unique, and in no case is the specifically-inserted HC found.**

The energy function and the perturbing moves we use are purely local affecting only four roads at a time. Yet, this energy function depending only on local information with some global jumps from the annealing yields effective results for a factorially growing problem. It seems that local information is essentially enough to solve a vast, complex global problem. It will be interesting to see at what size *N*, if any, our algorithm ceases to be effective. Or is global "knowledge" never needed? Why would a local algorithm work? For a few or many roads ($R << 0.5*N*\log_e N$ or $R >> 0.5*N*\log_e N$) the problem is easily solved[8]: For small *R* there are just not enough roads to find a tour, and for large R there is so much connectivity there almost always is a tour. The addition of the extra roads must provide our predominantly local algorithm with just enough connectivity or "information" to permit solution in the difficult intermediate case.

We do not have a proof that our algorithm will in all cases determine if a set of roads contains a HC, but practically the method is efficient, runs on a standard computer, and as far as can be tested is effective. Our results also bring up a new "constructive HCP" problem: Can a given HC be found in a set of roads? Finally, our approach and findings may be helpful in guiding proofs on the question of whether or not there is a polynomial time algorithm for NP-complete problems.

*This research used resources of the Argonne Leadership Computing Facility, which is a DOE Office of Science User Facility supported under Contract DE-AC02-06CH11357.*

# References

[1] Karp, R. Complexity of computer computations. R. Miller, J. Thatcher, J. Bohlinger, eds., *The IBM Research Symposia Series* (Springer US, 1972), pp. 85–103.

[2] Johnson, M.W., et al., Quantum annealing with manufactured spins. *Nature* **473,** 194-198 (2011).

[3] Rnnow, T. F. et al., Defining and detecting quantum speedup. *Science* **345,** 420-424 (2014).

[4] Kirkpatrick, S. Gelatt, Jr., C. D. & Vecchi, M. P. Optimization by simulated annealing. *Science* **220,** 671-680 (1983).

[5] Altschuler, E. L., Lades, M., & Stong, R. Finding Hamiltonian cycles. *Science* **273,** 413 (1996).

[6] Lin, S. & Kernighan, B. W. An effective heuristic algorithm for the traveling salesman problem. *Operations Research* **21,** 498-516 (1973).

[7] Metropolis, N. Rosenbluth, A. W., Rosenbluth, M. N., Teller, A. H. & Teller, E. Equation of state calculations by fast computing machines. *The Journal of Chemical Physics* **21,** 1087-1092 (1953).

[8] Bollobas, B. Fenner, T. & Frieze, A. An algorithm for finding Hamilton paths and cycles in random graphs *Combinatorica* **7,** 327-342 (1987).

[9] Kirkpatrick, S. & Selman, B. Critical behavior in the satisfiability of random Boolean expressions. *Science* **264,** 1297-1301 (1994).